\documentclass[aps,prl,reprint,showpacs,superscriptaddress]{revtex4-1}
\usepackage{graphicx} 
\usepackage{bm}

\begin{document}

\title{Spontaneous Distortion and Ferromagnetism Induced by Quantum-well States in Pd(100) Ultrathin Films}

\author{Shunsuke Sakuragi}
\email[e-mail: ]{sakuragi@az.appi.keio.ac.jp}
\affiliation{Department of Applied Physics and Physico-Informatics, Keio University, Hiyoshi, Yokohama 223-0061, Japan}

\author{Hiroo Tajiri}
\affiliation{ Japan Synchrotron Radiation Research Institute/SPring-8, Kouto, Sayo 679-5198, Japan}

\author{Hiroyuki Kageshima}
\affiliation{Interdisciplinary Graduate School of Science and Engineering, Shimane University, Nishikawatsu-cho, Matsue 690-8504, Japan}

\author{Tetsuya Sato}
\affiliation{Department of Applied Physics and Physico-Informatics, Keio University, Hiyoshi, Yokohama 223-0061, Japan}

\date{\today}

\begin{abstract}
We study the crystal structure of Pd(100) ultrathin films, which show ferromagnetism induced by the quantum confinement effect, using in-situ X-ray crystal truncation rod measurement and density functional calculation. The energy gain due to the appearance of ferromagnetism in Pd results in flatter and uniform film growth of ferromagnetic Pd films compared with paramagnetic Pd. In addition, ferromagnetic Pd films expand the lattice constant in order to suppress the increase in kinetic energy of electrons accompanied by the occurrence of exchange splitting. Although the traditional theory of magnetism in metals indicates that the increase in density of states that induces ferromagnetism (Stoner criterion), our present finding reveals a mechanism of modulation in the density of states via the appearance of ferromagnetism, i.e., the inverse mechanism of Stoner's theory.
\end{abstract}


\maketitle

Recently, the importance of magnetic materials in industry has increased due to the development of novel electronic devices based on the spin degree of freedom of electrons \cite{IgorRMP,BaderARCMP}. However, in bulk form elements, ferromagnetism appears in a small fraction of the 3$d$ transition metals, only Fe, Co and Ni. Over the past decade, the appearance of ferromagnetism in several metal particles, such as Pd, Pt and Au, has been reported \cite{TaniyamaEPL,ShinoharaPRL,SampedroPRL,YamamotoPRL}. 
The physical properties of metals are determined by electronic states near the Fermi energy $\epsilon _F$. 
Thus, it has been understood that the confinement of electrons, surface effect, and/or distortion in the crystal is accompanied by nano-scaling-induced ferromagnetism in the noble metals. 
Although the origin of ferromagnetism in nanoparticles remains to be solved, this phenomenon suggests the possibility of fabricating fascinating materials with novel properties not realized in bulk form.

Recently, it was shown that Pd(100) ultrathin films showed ferromagnetism in an oscillatory manner, dependent on film thickness \cite{niklassonPRB,mirbtPRB,hongPRB,SakuragiPRB}. 
From the standpoint of Stoner's theory for the appearance of ferromagnetism \cite{stoner}, 
\begin{equation}\label{stoner}
I D(\epsilon _F) > 1,
\end{equation}
where $I$ is the exchange integral and $D(\epsilon _F)$ is the density of states at the Fermi energy, the ferromagnetism in Pd(100) is interpreted based on the increase in $D(\epsilon _F)$ due to $d$-electron quantum-well states. 
The magnetization and Curie temperature of epitaxial Pd(100) ultrathin films on SrTiO$_3$(100) substrates are comparable to those of Ni \cite{SakuragiPRB}, where the magnetization is larger than one digit from the theoretical expectation \cite{hongPRB}. 
Elucidating the origin of the enhancement of ferromagnetic moment in Pd(100) would contribute to our understanding of the expression mechanism of ferromagnetism in transition metals, because the magnetism of this system can be systematically controlled through quantum-well states depending on film thickness, according to Stoner's theory. Because of its potential for the development of novel magnetic nanoscale materials whose magnetism can be controlled, this matter is of great technological interest \cite{ChibaNatMat,sunPRB,aiharaJAP, ShimizuPRL}.

In this Letter, we demonstrate the relationship between the crystal structure and magnetism in Pd(100) ultrathin films using surface X-ray diffraction and density functional calculations. The flat growth of Pd(100) films is attributed to the energy gain in the system accompanied by the appearance of ferromagnetism induced by quantum-well states, and 0.8\% lattice expansion was needed in order to suppress energy loss via the occurrence of exchange splitting. Our results suggest that the spontaneous stabilization in the magnetic state is brought about by lattice expansion via the appearance of ferromagnetism in the transition metal.

All experiments were performed at SPring-8 BL13XU \cite{sakataSRL}. 
We used atomically flat SrTiO$_3$(100) substrates treated with buffered hydrofluoric acid (SHINKOSHA Co., Ltd.) \cite{kawasakiScience}, and the following three-step growth method to prepare atomically flat Pd(100) ultrathin films \cite{wagnerJAP, SakuragiPRB, PhysProc}. 
First, we deposited 1/5 of total thickness of Pd film at 300 °C, and cooled it to room temperature. We then deposited 4/5 of total thickness of Pd. After deposition, post annealing was performed to $\sim$250 °C to improve the crystallinity. 
Finally, X-ray reflectivity and X-ray crystal truncation rod (CTR) scattering were measured in-situ, at room temperature using synchrotron X-rays at 15 keV. The pressure of the chamber was kept lower than $1\times10^{-7}$ Pa.

Fig. 1(a) shows X-ray reflectivities for several Pd films, whose thicknesses were evaluated from the fringe periods. It was found that visibility of reflectivity changes depended on film thickness. To make these trends clearer, we plotted ratios of peak-to-valley values in the fringes (i.e., peak/valley), as shown in Fig. 1(b). The peak/valley values oscillate as film thickness increases. Since visibility reflects film roughness in general, high peak/valley values indicate the film is flat. Here, we compare the present reflectivity results with the thickness-dependent magnetization in Pd(100) films \cite{SakuragiPRB} prepared under the same conditions as those shown in Fig. 1(b). It is apparent that both peak/valley values and the magnetic moment in Pd(100) show the same oscillatory behavior. This indicates that the improved uniformity of the film structure occurs due to the appearance of ferromagnetism.

\begin{figure}
\centering
\includegraphics[width=8.5cm]{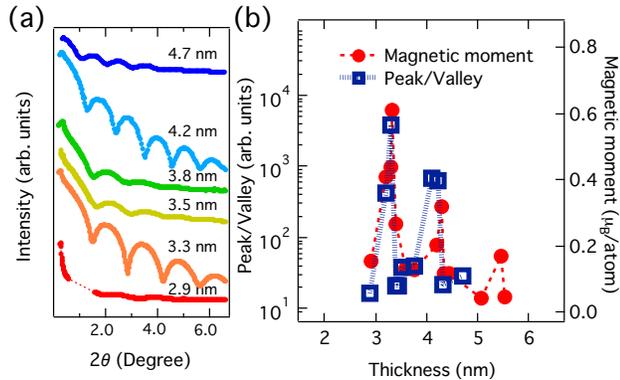}
\caption{\label{xrr} 
(a) X-ray reflectivities of Pd(100) ultrathin films of various thicknesses on SrTiO$_3$; (b) thickness dependence of peak/valley ratio of X-ray reflectivity. The thickness dependence of magnetic moments in Ref.  \cite{SakuragiPRB} is also shown for comparison.}
\end{figure}

The thickness dependence of 00 rod profiles of the X-ray CTR scatterings are shown in Fig. 2(a), which allows us to investigate the details of the film structures. 
The surface normal components $L$ of the Miller indices are expressed in the reciprocal lattice unit (r.l.u.) in the lattice constant of the SrTiO$_3$ substrate. 
All films exhibited Laue-function-like thickness fringes. The films of 3.3- and 4.2-nm thickness (i.e., the ferromagnetic films), where the peak/valley ratios of the reflectivities are highest, show regular forms of Laue-function oscillation. 
In other samples (namely near paramagnetic films), on the other hand, disturbed Laue-function oscillations were observed. These disturbances are attributed to spread in film thickness and lattice constants in the film, as revealed by the following analysis.

\begin{figure}
\centering
\includegraphics[width=8.5cm]{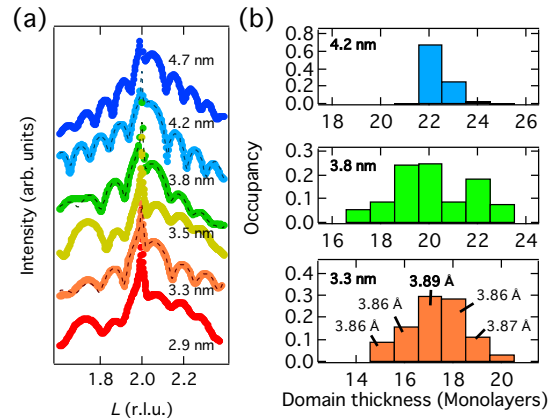}
\caption{\label{ctr} 
(a) Thickness-dependent 00 rod profiles of X-ray CTR scattering of Pd(100) ultrathin films on SrTiO$_3$. Dashed lines show the simulation data. (b) Thickness distribution of 4.2-, 3.8-, and 3.3-nm films. The lattice constant of each domain is shown for the 3.3-nm film. 
}
\end{figure}

For the quantitative discussion, we analyzed the profiles of X-ray CTR scatterings by least-squares fit. 
We assumed that the Pd(100) films had certain distributions in both film thickness and lattice constant on the Ti-O terminated SrTiO$_3$ substrate. The fitting results are shown in Fig. 2(a). 
The obtained film thickness distributions of samples are shown in Fig.2 (b); thicknesses estimated from reflectivity are 4.2 nm (ferromagnetic), 3.8 nm (paramagnetic), and 3.3 nm (ferromagnetic). 
A wide and anomalous thickness distribution, which indicates a disturbed film structure, was observed in the paramagnetic film with a thickness of 3.8 nm. 
On the other hand, ferromagnetic films had narrow thickness distributions signifying uniformity of film structure. In addition, the lattice constant of each domain in the ferromagnetic 3.3-nm sample was evaluated from the fitting, and the 17-monolayers thick domain, corresponding to the peak of the thickness distribution, showed $\sim$0.8\% lattice expansion, as may be seen in Fig. 2(b). This indicates that the appearance of ferromagnetism due to quantum-well states in Pd(100) ultrathin films promoted the atomic growth of flat film, which was accompanied by spontaneous lattice expansion.

This experiment shows the significant correlation between the appearance of ferromagnetism and the crystal structure of the film. 
In order to clarify the origin of this correlation, we performed a density function calculation. 
The PHASE/0 program \cite{PHASE} using the projector augmented wave-type (PAW) pseudopotential \cite{PAW} to the spin-polarized local density approximation reported by Perdew and Wang \cite{LDA} was used. 
The values of the lattice constant converge to 3.84 \AA for fcc bulk Pd, and we used this value for film-shaped Pd(100). 
To evaluate the magnetism of Pd(100) ultrathin films, a slab of Vacuum(2 monolayers)/Pd(N monolayers)/Vacuum(3 monolayers), $56 \times 56 \times 1$ $k$-points, and 36 Ry of cut-off energy was used. 
Based on this, we calculated the difference of total energies between paramagnetic and ferromagnetic states, where the spin polarization was fixed to a curtain value in freestanding Pd(100) \cite{aiharaJAP}. 
We calculated the magnetic states of freestanding Pd(100) from 2-23 monolayers, as shown in Fig. 3(a). Ferromagnetism appeared in Pd(100) depending on film thickness, in an oscillatory manner. The period of oscillation was consistent with previous calculations \cite{hongPRB} and experiments \cite{SakuragiPRB}, although the calculated ferromagnetic moment was smaller compared to previous findings. Despite the underestimation of the ferromagnetic moment of Pd, the present calculation is consistent with quantum-well induced ferromagnetism.

\begin{figure}
\centering
\includegraphics[width=8.5cm]{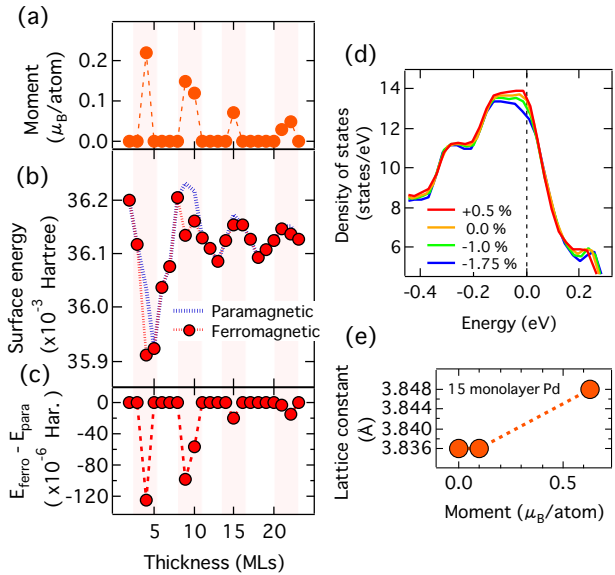}
\caption{\label{cal} 
(a) Calculation prediction of thickness-dependent magnetic moment in freestanding Pd(100) slabs. (b) Thickness-dependent surface energy of freestanding Pd(100) slabs with and without ferromagnetism. (c) The difference in surface energy between ferromagnetic states and paramagnetic states of Pd(100) slabs. (d) The total density of states of 10-monolayers Pd(100) slab near the Fermi energy as a function of out-of-plane lattice expansion, where 0.0 eV means the Fermi energy. (e) The out-of-plane lattice constant of the 15-monolayers Pd(100) slab with the appearance of quantum-well-induced ferromagnetism as a function of magnetic moment. 
}
\end{figure}

In order to investigate the stability of the film structure, the thickness dependence of the surface energy of freestanding Pd(100) was studied, as shown in Fig. 3(b), where the surface energy of paramagnetic Pd(100), which is calculated by assuming no spin polarization, is also shown for comparison. The difference of surface energies between ferromagnetic and paramagnetic films, shown in Fig. 3(c), indicates that there is a gain of surface energy due to the appearance of ferromagnetism. For ferromagnetic films with a large magnetic moment of Pd (monolayers 4 and 9), the surface energy decreased with the appearance of ferromagnetism compared to paramagnetic film of equivalent thickness. This indicates that the ferromagnetic domain is more suitable for growing, compared to paramagnetic domains of equivalent film thickness. This result can explain our experimental finding that ferromagnetic Pd films have a flat and uniform film structure.

Stoner's theory means that the ferromagnetic transition occurs if the energy gain due to exchange splitting can surpass the increase in the kinetic energy of electrons. This condition is satisfied when the system has a large $D(\epsilon _F)$. 
Pd has a sharp shape of density of states originating from 4$d$ bands near the Fermi energy, and thus its $D(\epsilon _F)$ is large compared to the other transition metals, even in bulk. When an increase in $D(\epsilon _F)$ occurs by quantum-well states, Stoner's criterion eq. (1) is satisfied in Pd(100), and there is an energy gain due to the appearance of spin polarization. 
Our results indicate that the energy gain due to spin polarization reduces the surface energy and makes ferromagnetic film flat. Crystal growth that depends on such a change in surface energy by nano-scaling is known as ``electronic growth'', which is the formation mechanism of flat metal films depending on film thickness \cite{ZhangPRL}. 
Our present finding, in other words, describes magnetic-induced electronic growth.

The experimentally observed spontaneous growth of flat film in ferromagnetic Pd(100) can be explained by the usual Stoner theory, as mentioned above, although the 0.8\% of lattice expansion due to the appearance of ferromagnetism is not explained. The lattice distortion brings changes to the $D(\epsilon _F)$ of transition metals as well as quantum-well states. The lattice expansion narrows the width of the band, and the shape of the total density of states becomes sharp. Fig. 3(d) shows total density of states near the Fermi energy of the Pd(100) ultrathin film, calculated with various out-of-plane lattice constants, where the 0.0\% means the lattice constant of bulk fcc Pd. The $D(\epsilon _F)$ increases with lattice expansion. This calculation result suggests that the $D(\epsilon _F)$ increases due to the appearance of ferromagnetism. Stoner's theory shows the mechanism by which the $D(\epsilon _F)$ determines the magnetic state. In contrast, our results indicate the mechanism of modulation in $D(\epsilon _F)$ due to spin polarization, i.e., the inverse mechanism of Stoner's theory.

Here, we calculate the magnetic moment as a function of the out-of-plane lattice constant of Pd(100) films in order to discuss the relationship between the appearance of ferromagnetism and spontaneous distortion [Fig. 3 (e)]. There is no change in the lattice constant of Pd between nonmagnetic states and weak ferromagnetic states with a spontaneous moment of 0.1 $\mu_B$/atom. On the other hand, the 0.3\% lattice expansion is observed at the spontaneous moment of the 0.6 $\mu_B$/atom, which is the experimentally obtained value of the magnetic moment. This result supports the existence of a mechanism by which the system is stabilized through the spontaneous change in electronic states accompanied by the occurrence of the huge exchange splitting; that is, the inverse effect of Stoner's theory.

When the exchange splitting occurs, the kinetic energy of the electrons with the majority spin increases. The band narrowing, caused by lattice expansion, reduces the number of electrons with high kinetic energy, and thus suppresses the increase of total energy in the system by exchange splitting. Therefore, this mechanism explains the spontaneous lattice expansion in Pd(100) in transitioning to a ferromagnetic state, as shown schematically in Fig. 4.

\begin{figure}
\centering
\includegraphics[width=8.5cm]{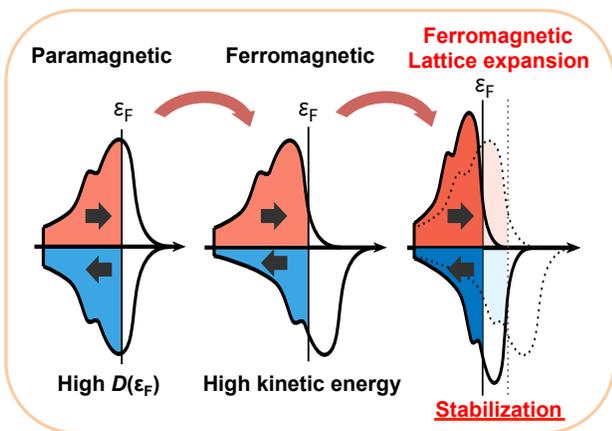}
\caption{\label{schematic} 
A schematic image of spontaneous lattice distortion induced by the appearance of ferromagnetism. 
}
\end{figure}

The Stoner criterion means that the increase in $D(\epsilon _F)$ stabilizes the ferromagnetic state more than the paramagnetic state. Therefore, spontaneous lattice expansion in ferromagnetic Pd(100) ultrathin films can occur to stabilize ferromagnetic states of Pd.

The stabilization of the magnetic states by spontaneous lattice distortion was previously predicted for Fe, which is a typical ferromagnetic transition metal \cite{ShigaJPSJ,EkmanPRB,MankovskyPRB}. In previous density functional calculations, it was reported that the 2\% lattice expansion in bulk bcc Fe via the appearance of ferromagnetism altered electronic states to spontaneously stabilize the magnetic states. In our present study, this theoretical prediction was verified by systematic experiments using quantum-well induced ferromagnetism in Pd.

In conclusion, we investigated the thickness-dependent structural change in Pd(100) ultrathin films accompanied by a change in magnetic state in terms of quantum-well induced ferromagnetism. When spontaneous magnetization is induced in Pd(100) by quantum-well states, stabilization of film structure by reduced surface energy was observed. In addition, a reduction of the total energy of the system by spontaneous lattice distortion occurred, accompanied by the appearance of ferromagnetism. Our findings suggest a mechanism for determining the magnetic states in transition metals by the mutual relations between magnetism, quantum-well states, and lattice distortion. This mechanism can be extended to other magnetic materials, opening the possibility of tailoring magnetic states and/or magnetization by appropriate structural electronic engineering.

We thank Y. Watanabe for technical support with respect to the X-ray diffraction measurements and R. Itotani, S. Urasaki, and K. Okada for fruitful advice on theoretical calculation. 
The synchrotron radiation experiments were performed at the BL13XU of SPring-8 with the approval of the Japan Synchrotron Radiation Research Institute (JASRI) (Proposal No. 2014A1675, 2015A1775, 2015B1689). 
This work was supported by JSPS KAKENHI Grant Numbers \#15J00298 and 15H01998. 
One of the authors (S.S.) also acknowledges a fellowship from the JSPS.

\bibliography{ref}
\end{document}